\documentclass[12pt]{iopart}
\pdfoutput=1
\usepackage{graphicx}
\usepackage{color}
\begin{document}

\title[What-Happens-Next-Machine]
{Kermit's What-Happens-Next-Machine Reloaded}

\author{U Schr\"oter}
\address{University of Konstanz, FB Physik,
Universit\"atsstra\ss e 10, 78464 Konstanz, Germany}

\ead{Ursula Schroeter@uni-konstanz.de}

\begin{abstract}
If Kermit's What-Happens-Next-Machine had functioned, you would not have
seen much, because it would have gone too quickly. In this article it is
shown that putting up and solving the equations of motion of a seemingly
simple mechanical apparatus presents a challenging problem. The simulation
can, however, be quite instructive and also entertaining.
\end{abstract}


\section{Introduction}

\phantom{m}\hspace{-1.5cm}
\parbox{8cm}{\includegraphics[width=7cm,angle=270]{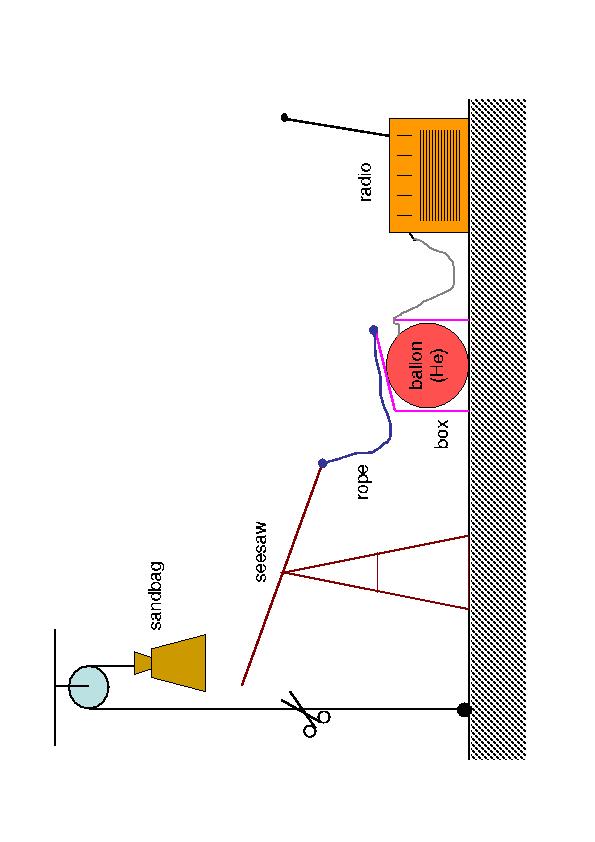}

\vspace{-7.3cm}

\phantom{m}\hspace{4.5cm}\includegraphics[width=3cm,angle=270]{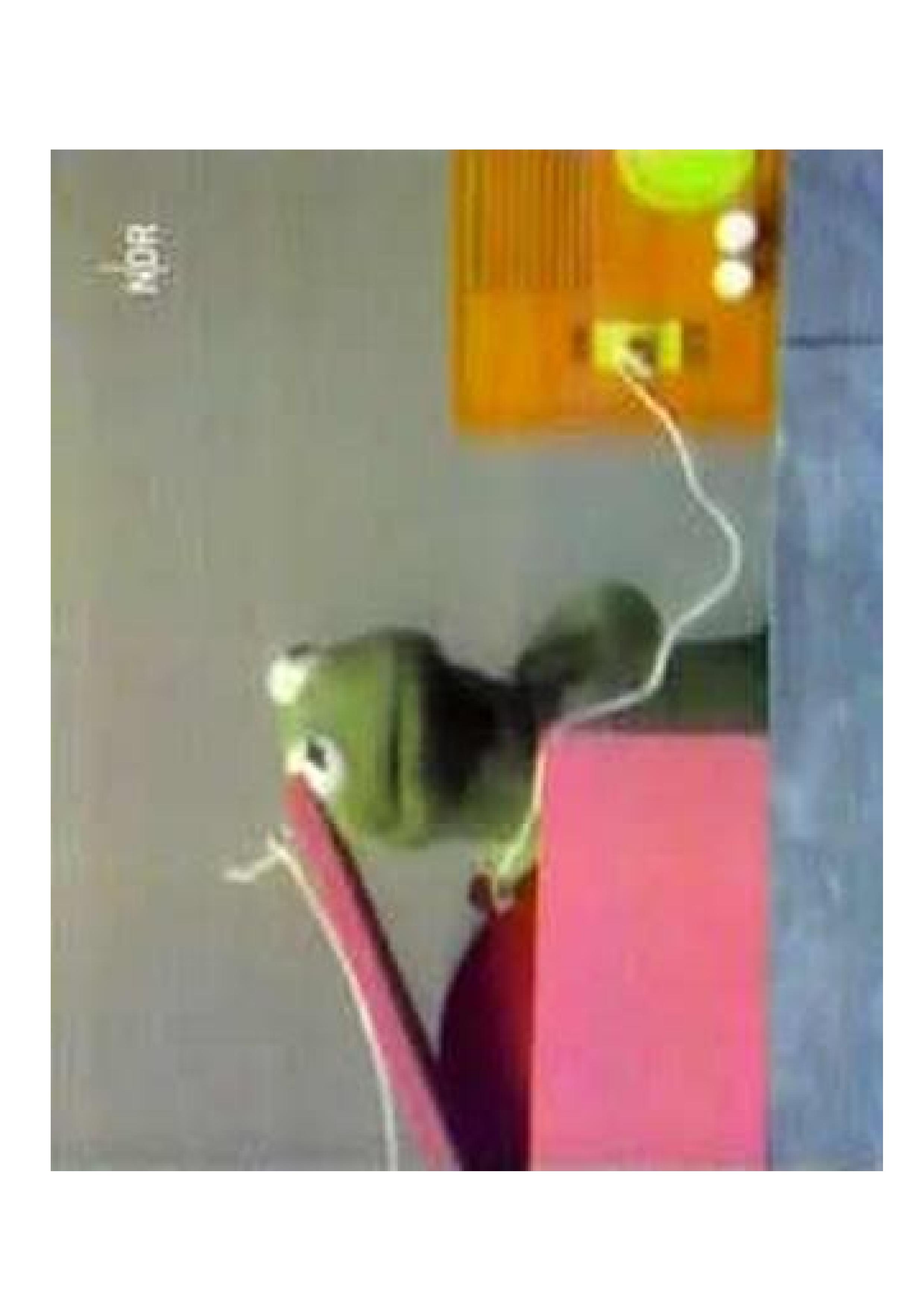}

\vspace{3cm}

\phantom{nix}}
\parbox{1cm}{\phantom{nix}}
\parbox{8cm}{\includegraphics[width=7cm,angle=270]{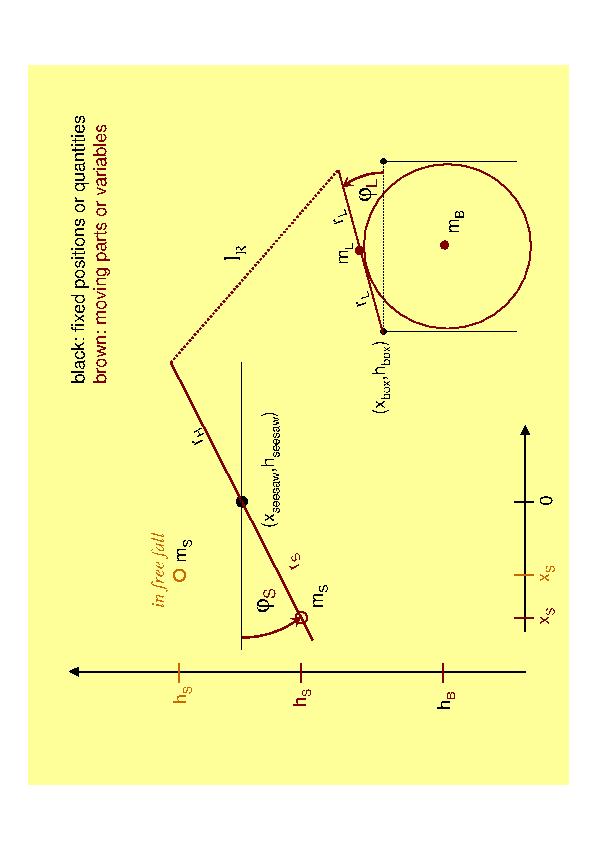}
}

\phantom{m}\hspace{-1cm}
\parbox{8cm}{{\small {\bf Fig.1} Scheme of the What-Happens-Next-Maschine.
Inset: Kermit the Frog in the scene.}}
\parbox{1cm}{\phantom{nix}}
\parbox{8cm}{{\small {\bf Fig.2} Geometry of the seesaw and the box's lid.
Definition of positions, distances and angles.}}

\vspace{0.5cm}

In the children's television series {\it Sesame Street} Kermit the Frog,
a muppet created by Jim Henson, presented his What-Happens-Next-Machine.
Although he proclaimed having constructed it to avoid walking from here to
there to turn on his radio, the purpose of the scene, of course, lies in
conveying to the spectator the concept of a chain reaction. More spectacular
chain reactions are now often demonstrated in science museums \cite{nemo}
and TV shows \cite{domino}. The What-Happens-Next-Machine, however,
is a - if not {\it the} - historical, famous or notorious example. The setup
is meant to function as follows: A string being cut, a sandbag hung from a
pulley drops onto a seesaw. To the other end of the seesaw a rope is tied
which lifts the lid of a box containing a balloon, preferably filled with
helium. From the balloon a string had already been trailing out of the box and
is attached to the switch of a radio. So as the lid opens and the balloon
rises, at the point the strings is pulled straight up the radio is turned on.
But, fitting the typical irony of Sesame Street, even though Kermit tries to
save whatever possible, nearly every step - except for the balloon floating
up (with some delay) - is malfunctioning \cite{youtube}. Thus, albeit for
the good entertainment the flawed performance provides, unfortunately the
spectator does not get to see the before described and intended chain reaction.
Without malfunctions, how fast would it go?

\phantom{m}\hspace{-2.5cm}
\parbox{8cm}{\includegraphics[width=7cm,angle=270]{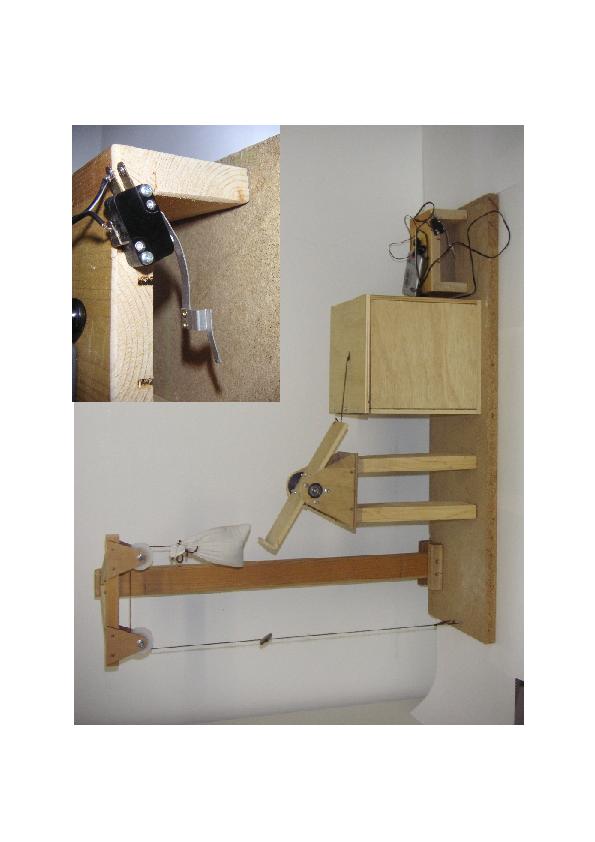}}
\parbox{1cm}{\phantom{nix}}
\parbox{8cm}{\includegraphics[width=7cm,angle=270]{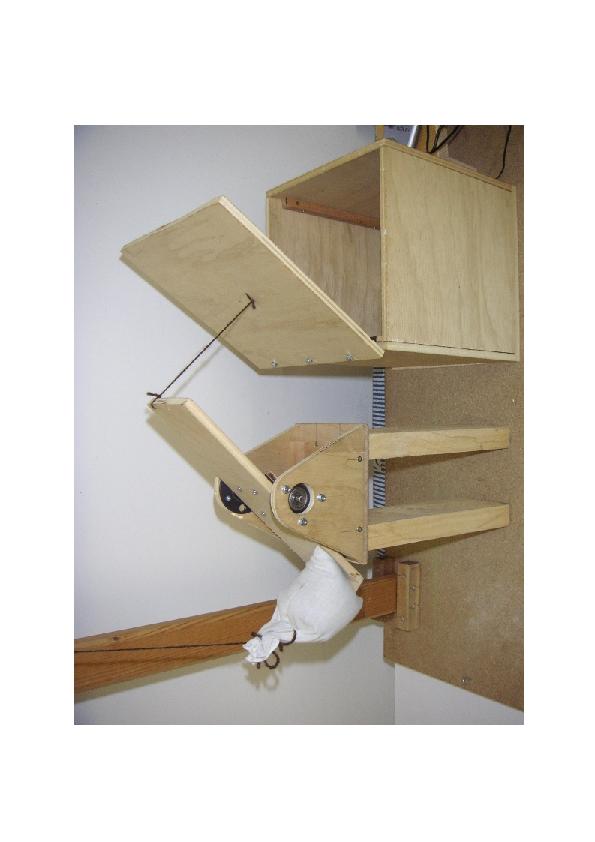}}

\phantom{m}\hspace{-1cm}
\parbox{8cm}{{\small {\bf Fig.3} What-Happens-Next-Machine built by students
in Konstanz 2009. Slightly diffent from the original and theoretically discussed
version, the rope is attached at the middle of the box's lid and the seesaw's
little edge will prevent the sandbag from sliding. The inset shows
a rocker switch a straightening ballon string can actually operate.)
}}
\parbox{1cm}{\phantom{nix}}
\parbox{8cm}{{\small {\bf Fig.4} Here the sandbag has fallen, turned over the
seesaw and opened the lid. In this foto taken during setup there is
no balloon yet. The seesaw need not pull the lid up to the vertical. The balloon
can be smaller than the box and come out sideways. Anyway, if the
lid is pulled strongly enough at the beginning, it has enough momentum to
move on up to vertical.}}

\vspace{0.5cm}

While it can also be a nice task for students to build and run the machine,
this article focusses on calculating the motion. Referring to every-day life
phenomena or using toys \cite{toys} can make physics courses and exercises
more appealing. The difficulty often is that one physical concept
in its pure form is insufficient to explain the behaviour. A calculation of
the What-Happens-Next-Machine as well has to begin by making appropriate
idealizations. It can, however, provide a challenging exercise for second-year
students of physics. In analytical mechanics textbook examples may fall short
of actually conveying the virtue of the Langrange formalism, because the posed
problems of calculating something's motion can just as well be solved by
Newtonian concepts, evaluating forces and using energy and momentum conservation.
In the setup considered here in the connection of the seesaw to the also rotating lid
by the rope an unexpected complexity is concealed. Finally, numerics will be
used making the theoretical problem also a programming exercise.

\section{Lagrangian}

The setup is drawn in Fig.1 and a close-up with labels is given in Fig.2.
We need:

{\small

\begin{tabular}{ll}
\multicolumn{2}{l}{fixed quantities:} \\
$g$ : & constant of gravitation, $9.81\; m\; s^{-2}$ \\
$m_S$ : & mass of the sandbag (considered a point mass) \\
$r_H$ : & length of the right half of the seesaw \\
$l_R$ : & length of the rope (straightened) \\
$r_L$ : & half edge of the box, and thus also of the lid (assumed square) \\
$m_L$ : & mass of the box's lid \\
$I_L$ : & moment of inertia of the box's lid with respect to its hinge;
$I_L=m_L\cdot (2r_L)^2/3$ \\
$r_B$ : & radius of the balloon (assumed a sphere); we set $r_B=r_L$ \\
$\rho_{Air}$ : & density of air, 1.29 kg m$^{-3}$
\ \ \ $\rho_{He}$: \ density of helium, 0.18 kg m$^{-3}$ \\
\multicolumn{2}{l}{variables:} \\
$x_S$ : & horizontal position of the sandbag counted from the middle of the seesaw \\
$h_S$ : & vertical position (height) of the sandbag also counted from the middle of the seesaw \\
$r_S$ : & distance of the sandbag from the middle of the seesaw when the two are in contact \\
$\varphi_S$ : & tilt angle of the seesaw, zero=horizontal,
pos.=left side is down, neg.=left side is up \\
$\varphi_L$ : & opening angle of the lid, counted positive when opened from horizontal \\
$h_B$ : & height of the balloon (center) \\ 
\end{tabular}

\vspace{0.3cm}

The following energies are involved:

\begin{tabular}{lll}
\underline{component} \phantom{nixx} & \underline{potential} & \underline{kinetic} \\
sandbag & $m_S\; g\; h_S$  & $\frac{m_S}{2}\; (\dot{x}_S^2+\dot{h}_S^2)$ \\
lid & $m_L\; g\; r_L\; \sin\varphi_L$ & $\frac{I_L}{2}\; \dot{\varphi}_L^2$ \\
balloon & $\frac{4}{3}\pi \; r_B^3\; (\rho_{He}-\rho_{Air})\; g\; h_B$
\phantom{nix} &
$\frac{1}{2}\; \frac{4}{3}\pi \; r_B^3\; \rho_{He} \; \dot{h}_B^2$ \\
\end{tabular}
}

\vspace{0.3cm}

$h_S$=$-r_S\sin \varphi_S$ if the sandbag is on the seesaw.
The kinetic energy of the sandbag can be rewritten as
$\frac{m_S}{2}(\dot{r}_S^2+r_S^2 \dot{\varphi}_S^2)$
if the sandbag and the seesaw are moving together.
The potential energy of the balloon accounts for buoyancy.
The seesaw is regarded as massless (that is negligible mass as compared to
the sandbag). The box is fixed to the ground, only the lid moves. The rope
is assumed massless and the mass of the balloon's rubber skin is neglected,
too \cite{praxis}.
There is no friction in the hinges of the seesaw and the lid
and as well no sliding friction is considered should the sandbag slide down
the seesaw. The Lagrangian \cite{anamech} (for the sandbag in contact with the seesaw) is
therefore put up as:
\begin{eqnarray}
\mathcal{L}&=&\frac{m_S}{2}(\dot{r}_S^2+r_S^2\dot{\varphi}_S^2)+
m_Sgr_S\sin\varphi_S+\frac{I_L}{2}\dot{\varphi}_L^2-m_Lgr_L\sin\varphi_L \\
&& +\frac{1}{2}\frac{4}{3}\pi r_B^3\rho_{He}\dot{h}_B^2-
\frac{4}{3}\pi r_B^3(\rho_{He}-\rho_{Air})gh_B
\nonumber\end{eqnarray}
In the following we shall, however, consider $\mathcal{L}$ without the balloon's
contributions. To neglect the pressure the balloon due to buoyancy exerts
from below onto the lid, seems a realistic assumption. The balloon will
therefore only follow the motion of the lid if it can rise fast enough.
We further assume that the balloon does not bounce off downwards when it
hits the lid, but that its motion is completely damped out in this case.
Letting aside the balloon, following the Lagrange formalism we deduce:
\begin{eqnarray}
(\frac{\partial\mathcal{L}}{\partial r_S}-\frac{d}{dt}
\frac{\partial\mathcal{L}}{\partial\dot{r}_S})\delta r_S +
(\frac{\partial\mathcal{L}}{\partial \varphi_S}
-\frac{d}{dt}\frac{\partial\mathcal{L}}{\partial\dot{\varphi}_S})
\delta\varphi_S
+ (\frac{\partial\mathcal{L}}{\partial\varphi_L}
-\frac{d}{dt}\frac{\partial\mathcal{L}}{\partial
\dot{\varphi}_L})\delta\varphi_L
= \nonumber\\
(m_Sr_S\dot{\varphi}_S^2+m_Sg\sin\varphi_S-m_S\ddot{r}_S)\delta r_S +
(m_Sgr_S\cos \varphi_S - 2m_Sr_S\dot{r}_S\dot{\varphi}_S - \nonumber \\
m_Sr_S^2\ddot{\varphi}_S)\delta\varphi_S -
(m_Lgr_L\cos{\varphi}_L+I_L\ddot{\varphi}_L)\delta\varphi_L
= 0 \end{eqnarray}
$t$ is time. $r_S$ is an independent variable, whence we get
\begin{equation}
\ddot{r}_S=g\; \sin\varphi_S +r_S\; \dot{\varphi}_S^2
\end{equation}
as a first differential equation or equation of motion. With the rope straight,
$\varphi_S$ and $\varphi_L$ are not independent. We choose to take $\varphi_L$
as a function of $\varphi_S$. The geometric relation is given in an attachment.
Then $\delta\varphi_L=\frac{d\varphi_L}{d\varphi_S}\delta\varphi_S$,
$\dot{\varphi}_L=\frac{d\varphi_L}{d\varphi_S}\dot{\varphi}_S$ and
$\ddot{\varphi}_L=\frac{d^2\varphi_L}{d\varphi_S^2}\dot{\varphi}_S^2+
\frac{d\varphi_L}{d\varphi_S}\ddot{\varphi}_S$.
The derivatives $\frac{d\varphi_L}{d\varphi_S}$ and
$\frac{d^2\varphi_L}{d\varphi_S^2}$
are evaluated numerically in an actual calculation. Thus
a second equation of motion follows from (2):
\begin{equation}
\ddot{\varphi}_S=\frac{m_Sgr_S\cos\varphi_S-2m_Sr_S\dot{r}_S\dot{\varphi}_S-
(m_Lgr_L\cos\varphi_L+I_L\frac{d^2\varphi_L}{d\varphi_S^2}\dot{\varphi}_S^2)
\frac{d\varphi_L}{d\varphi_S}}{m_Sr_S^2+I_L(\frac{d\varphi_L}{d\varphi_S})^2}
\end{equation}
The functionality $\varphi_L(\varphi_S)$ is a constraint in the sense of the
Lagrange formalism. It takes the form of an equality if the rope is straight.
A non-holonomous constraint, nevertheless, is also always present if the rope
is loose, namely that the right endpoint of the seesaw and the right edge of
the lid cannot be more than a distance of $l_R$ apart. Even if here we do not
handle the balloon as part of one complete Lagrangian, there is an also
non-holonomous constraint on its motion. It cannot exceed a certain height
determined by the lid, whereas it may well stay below that limit. Geometric
considerations concerning this point are given in an attachment. Furthermore,
not including the balloon pressing against the lid from underneath into the
equations of motion, care must be taken not to let the lid drop below its
initial position on top of the balloon. The integration of (3) and (4) is
performed numerically, in the most simple manner in the attached program.
More advanced methods like, for example, the Runge-Kutta scheme, could of
course be employed.

\section{Impact}

Calculating the free fall of the sandbag as well as the time and position
of its hitting the seesaw is trivial. At the beginning the seesaw can be
at rest at any angle the length of the rope allows. As long as or if the rope
is not straight our massless seesaw will just give way to the free falling
sandbag. However, the instant the rope straightens or is already straight
when the sandbag hits the seesaw, because of suddenly the weight of the lid
pulling there will be a discontinuity in the velocities, which must be
determined apart from equations of motion like (3) and (4).
This turned out to be the most difficult, and to my opinion still somewhat
ambiguous aspect of the problem. For reasons
explained later the more general case is investigated that the lid can
already be moving when the rope straightens or the sandbag hits the seesaw.
Furthermore can the sandbag also have a horizontal velocity in its free fall. 
As there is nothing to stop it, the velocity of the sandbag along the seesaw
$v_{\parallel}$ is conserved and assigned to $\dot{r}_S$ at impact.
\begin{equation}
\dot{r}_S=-\dot{x}_S\cos\varphi_S-\dot{h}_S\sin\varphi_S
\end{equation}
But what about the initial angular velocity of the seesaw immediately after
impact? We shall assume an inelastic collision insofar as the sandbag never
bounces off upwards from the seesaw. Immediately after impact the seesaw and
the sandbag move together (later it is checked whether the vertical velocity
from the turning and centrifugal motion would exceed free fall and eventually
the sandbag is allowed to loose touch with the seesaw). Both kinetic energy
and angular momentum cannot be conserved in the collision. Drawing the analogy
to the one-dimensional inelastic collision where the two bodies stick together
afterwards, momentum is conserved, but part of the kinetic energy goes into
deformation and heat, the logical decision in our case is to require conservation
of angular momentum. However, as the sandbag on the seesaw and the lid do not
turn around the same center, the definition of angular momentum is not necessarily
clear here. It further only imports at the singular moment of impact. (The
motion afterwards of all parts together as calculated from the Lagrangian obeys energy
conservation.)

Our treatment of the impact is therefore based on the following
very elementary considerations
\footnote{Nevertheless, conserving the velocity of the sandbag along the seesaw
and for the rotations angular momentum in the form $m_Sr_S^2\dot{\varphi}_S+
I_L\dot{\varphi}_L$ does {\it not} produce spurious results
($r_S$ and $\dot{\varphi}_S$ before the collision just
describe the sandbag at the end of its free fall in polar coordinates,
$\dot{\varphi}_L$ before is known, afterwards
$\dot{\varphi}_L = \dot{\varphi}_L(\varphi_S,\dot{\varphi}_S)$, thus only
$\dot{\varphi}_S$ needs to be determined). 
Neither does conserving kinetic energy of rotational motion $\frac{m_S}{2}
r_S^2\dot{\varphi}_S^2+\frac{I_L}{2}\dot{\varphi}_L^2$ produce any qualitatively
or obviously impossible behaviour (with in general the rope at oblique angles
to the seesaw and the lid, forces perpendicular to one can well be along the
other, and thus conservation of the sum of angular momenta of each around
their hinge need not necessarily be expected, which would leave energy as a
quantity eventually to be conserved).
Attempting to evaluate velocities after the collision by normal momentum
conservation together with the constraint that afterwards the sandbag on the
seesaw and the lid at least momentarily move together
does lead to contradictions, even if
the angles $\varphi_S(t)$ and $\varphi_L(t)$ do not show them. There is more
kinetic energy after the collision than before. And this does produce cases
where the sandbag bounces off the seesaw with an upward velocity component.}:
The instant the sandbag hits and/or the rope straightens, the pulling lid acts
like a hypothetical mass on the seesaw at distance $r_H$ from the center. The
angular momentum the lid has with respect to its hinge can instead be taken as
that of one third of the mass sitting at its end. Therefore decompose
$I_L\dot{\varphi}_{L,before}$ into a lever arm $2r_L$ and momentum perpendicular
to it $p_{\perp,L}=\frac{I_L\dot{\varphi}_{L,before}}{2r_L}$. However, now in order to
find the part having an effect at the other end of the rope tied to the seesaw,
the momentum has to viewed as made up of two in general non-perpendicular
components along the lid and along the rope, the latter being
$p_{R,L}=\frac{I_L\dot{\varphi}_{L,before}}{2r_L\sin\alpha}$. How to evaluate the
angle $\alpha$ is also given in an attachment. At the seesaw this momentum
translated along the rope with lever arm $r_H$ makes up an angular momentum
$\frac{I_L\dot{\varphi}_{L,before}r_H\sin(\alpha+\varphi_S-\varphi_L)}{
2r_L\sin\alpha}$ (Fig.5). After the collision the change of $\varphi_L$ is determined
by that of $\varphi_S$, nevertheless, the ''transfer coefficient'' for angular
momentum through the rope remains the factor $\frac{r_H\sin(\alpha+\varphi_S-
\varphi_L)}{2r_L\sin\alpha}$. With $v_{\perp}=\dot{x}_S\sin\varphi_S-
\dot{h}_S\cos\varphi_S$ the velocity of the sandbag perpendicular to the
seesaw before, to solve for $\dot{\varphi}_S$ immediately after impact we
have the following equation:
\begin{eqnarray}
m_Sr_Sv_{\perp}+\frac{I_L\dot{\varphi}_{L,before}r_H\sin(\alpha+\varphi_S-
\varphi_L)}{2r_L\sin\alpha}=\nonumber \\ \phantom{etwasnachrechts}
(m_Sr_S^2+\frac{I_Lr_H\sin(\alpha+\varphi_S-
\varphi_L)\frac{d\varphi_L}{d\varphi_S}}{2r_L\sin\alpha})\dot{\varphi}_S
\end{eqnarray}
The first time the sandbag hits when it comes from vertical free fall
and the lid is still at rest, we have $\dot{x}_S=0$ and $\dot{\varphi}_{
L,before}=0$ at impact. Of course, the argument could have been led the other
way round, first replacing $m_S$ at $r_S$ by some mass turning at $r_H$ on
the seesaw, translating a momentum component down the rope to the lid and
establishing conservation of angular momentum around the hinge of the lid.
The results for $\dot{\varphi}_S$ and $\dot{\varphi}_L$ after the collision
would have been the same.

\phantom{m}\hspace{-1.5cm}\parbox{8cm}{
\includegraphics[width=6.5cm,angle=270]{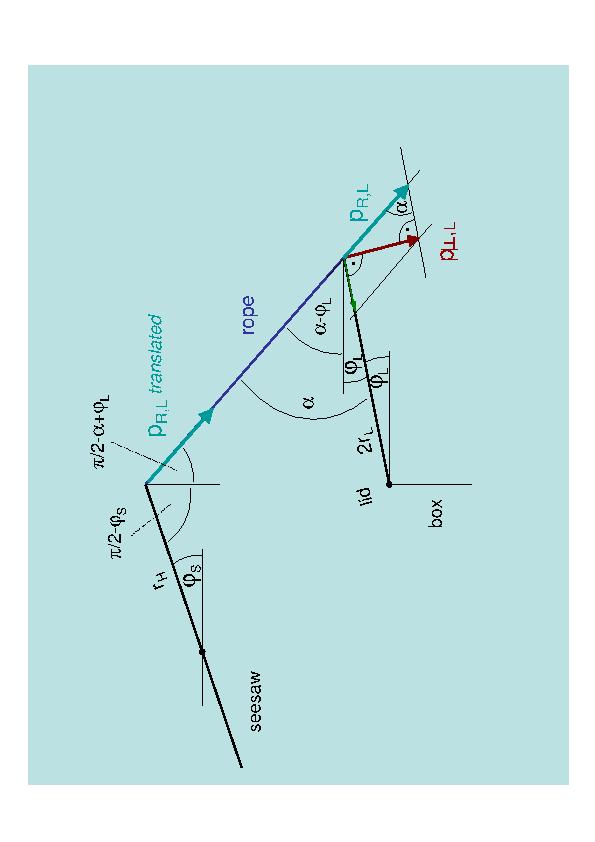}
}\parbox{1cm}{\phantom{nix}}\parbox{8cm}{
\includegraphics[width=6.5cm,angle=270]{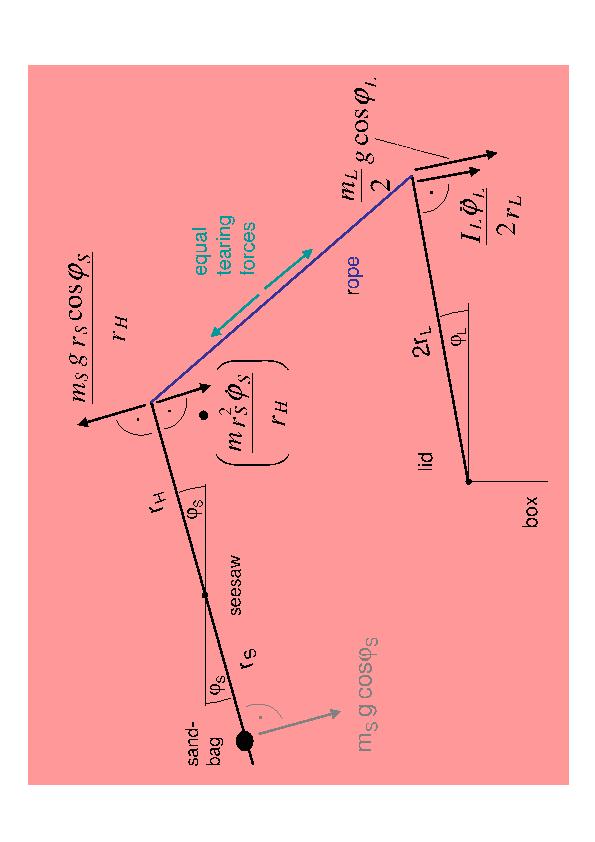}}

\phantom{m}\hspace{-1cm}\parbox{8cm}{
{\small {\bf Fig.5} Determining angles and evaluating the momentum component
of the lid along the rope.}
}\parbox{1cm}{\phantom{nix}}\parbox{8cm}{
{\small {\bf Fig.6} Gravitational and acceleration forces and tearing forces
on the rope.}}

\section{Torques, forces and energy}

In this section it shall be sought whether the equation of motion could have
been deduced by more elementary means than the Lagrange formalism. With the
sandbag on the seesaw the sum of kinetic and potential energies is
\begin{equation}
E=\frac{m_S}{2}(\dot{r}_S^2+r_S^2\dot{\varphi}_S^2)-m_Sgr_S\sin\varphi_S+
\frac{I_L}{2}\dot{\varphi}_L^2+m_Lgr_L\sin\varphi_L
\end{equation}
Requiring energy conservation leads to:
\begin{eqnarray}
0&=&\frac{dE}{dt}=m_S\dot{r}_S\ddot{r}_S-m_Sr_S\dot{r}_S\dot{\varphi}_S^2
-m_Sg\dot{r}_S\sin{\varphi}_S+2m_Sr_S\dot{r}_S\dot{\varphi}_S^2
\nonumber \\ &&
+m_Sr_S^2\dot{\varphi}_S\ddot{\varphi}_S-m_Sgr_S\dot{\varphi}_S\cos\varphi_S
+I_L\dot{\varphi}_L\ddot{\varphi}_L+m_Lgr_L\dot{\varphi}_L\cos{\varphi}_L
\end{eqnarray}
where the term $m_Sr_S\dot{r}_S\dot{\varphi}_S^2$ has been split into a negative
and twice the positive contribution. Now using the knowledge that the radial
motion of the sandbag will follow the respective component of the gravitational
force and the centrifugal force, futher demand that the first three terms in
(8) and the rest be zero each on its own. Dividing by $m_S\dot{r}_S$ or
$\dot{\varphi}_S$, respectively, and expressing $\dot{\varphi_L}$ and
$\ddot{\varphi}_L$ by derivatives and $\dot{\varphi}_S$ and $\ddot{\varphi}_S$
as before for the motion with the rope straight, this will reproduce (3) and (4).

However, as likely if not more likely than by putting up energy conservation,
one would approach a mechanics problem with levers and ropes by looking at
forces and torques. The following considerations are still restricted to the
motion when the sandbag is in touch with the seesaw and the rope is straight.
As explained earlier, the first equation of motion (3) for $\ddot{r}_S$ is
rather obviously written down. The torques determining the angular accelerations
$\ddot{\varphi}_S$ and $\ddot{\varphi}_L$ are more difficult to find. The time
derivative of the sandbag's angular momentum $m_Sr_S^2\dot{\varphi}_S$ is
$m_Sr_S^2\ddot{\varphi}_S+2m_Sr_S\dot{r}_S\dot{\varphi}_S$. Like in the last
section the change in angular momentum of the lid is taken into account as seen
at the other end of the rope. Therefore with respect to the seesaw's center
this is $I_L\ddot{\varphi}_L\frac{r_H\sin(\alpha+\varphi_S-\varphi_L)}{
2r_L\sin\alpha}$. The component of the gravitational force on the sandbag
which leads to angular acceleration is $m_Sgr_S\cos\varphi_S$. For the torque
$m_L$'s gravitation exerts on the lid replace the the lid by half the mass
$\frac{m_L}{2}$ at the end of an otherwise massless lid. As force component
perpendicular to the lid there you have $\frac{m_L}{2}g\cos\varphi_L$;
multiply by $2r_L$ to get the torque. This torque translates into an effective
one $\frac{m_L}{2}g\cos{\varphi}_L\cdot 2r_L\cdot\frac{r_H\sin(\alpha+\varphi_S-
\varphi_L)}{2r_L\sin\alpha}$ at the other end of the rope up at the seesaw,
which is, of course, opposed to the sandbag's. The seesaw and the lid are
literally tied together and therefore the available torques together have to
provide the changes in angular momentum taken together:
\begin{eqnarray}
m_Sr_S^2\ddot{\varphi}_S+2m_Sr_S\dot{r}_S\dot{\varphi}_S+I_L\ddot{\varphi}_L\cdot \frac{r_H\sin(
\alpha+\varphi_S-\varphi_L)}{2r_L\sin\alpha}= \nonumber \\
\phantom{etwasPlatz} m_Sgr_S\cos{\varphi}_S-\frac{m_L}{2}g\cos\varphi_L\cdot
2r_L \cdot \frac{r_H\sin(\alpha+\varphi_S-\varphi_L)}{2r_L\sin\alpha}
\end{eqnarray}
We checked numerically that $\frac{d\varphi_L}{d\varphi_S}=\frac{r_H\sin(
\alpha+\varphi_S-\varphi_L)}{2r_L\sin\alpha}$. And when again $\ddot{\varphi}_L
=\frac{d^2\varphi_L}{d\varphi_S^2}\dot{\varphi}_S^2+\frac{d\varphi_L}{
d\varphi_S}\ddot{\varphi}_S$ is implied,
(9) becomes identical to (4). Setting up the equation of motion for
$\ddot{\varphi}_S$ from forces and torques one is likely to forget the term
$2m_Sr_S\dot{r}_S\dot{\varphi}_S$, which looks like a Coriolis force \cite{alonso},
because (3) already seems to cover effects associated to the radial motion.
Instead of projecting the lid's change in angular momentum into an acceleration
of rotational motion of the seesaw, one could decompose forces that lead to
the respective torques at the right end of the seesaw and at the end of the lid
into parts along the seesaw or lid and along the rope and say that the tearing
force on the rope is equal in both directions (Fig.6). The acceleration forces are
against the sense of the angles, but an equivalent force on the right side of
the seesaw to the weight of the sandbag on the left has to point upwards.
Further $r_S/r_H$ is for changing the lever arm of the sandbag from $r_S$ to $r_H$
and $\sin(\alpha+\varphi_S-\varphi_L)$ and $\sin\alpha$ are the already known
geometrical factors from the parallelograms of forces (or momenta) to get
parts along the rope.
\begin{equation}
\frac{1}{\sin\alpha}(\frac{m_L}{2}g\cos\varphi_L+\frac{I_L\ddot{\varphi}_L}{
2r_L})=\frac{m_Sgr_S\cos\varphi_S-m_Sr_S^2\ddot{\varphi}_S-2m_Sr_S\dot{r}_S
\dot{\varphi}_S}{r_H\sin(\alpha+\varphi_S-\varphi_L)}
\end{equation}
(10) is the same as (9).

\section{Cases}

\setcounter{figure}{6}

\begin{figure}\begin{center}
\includegraphics[width=12cm]{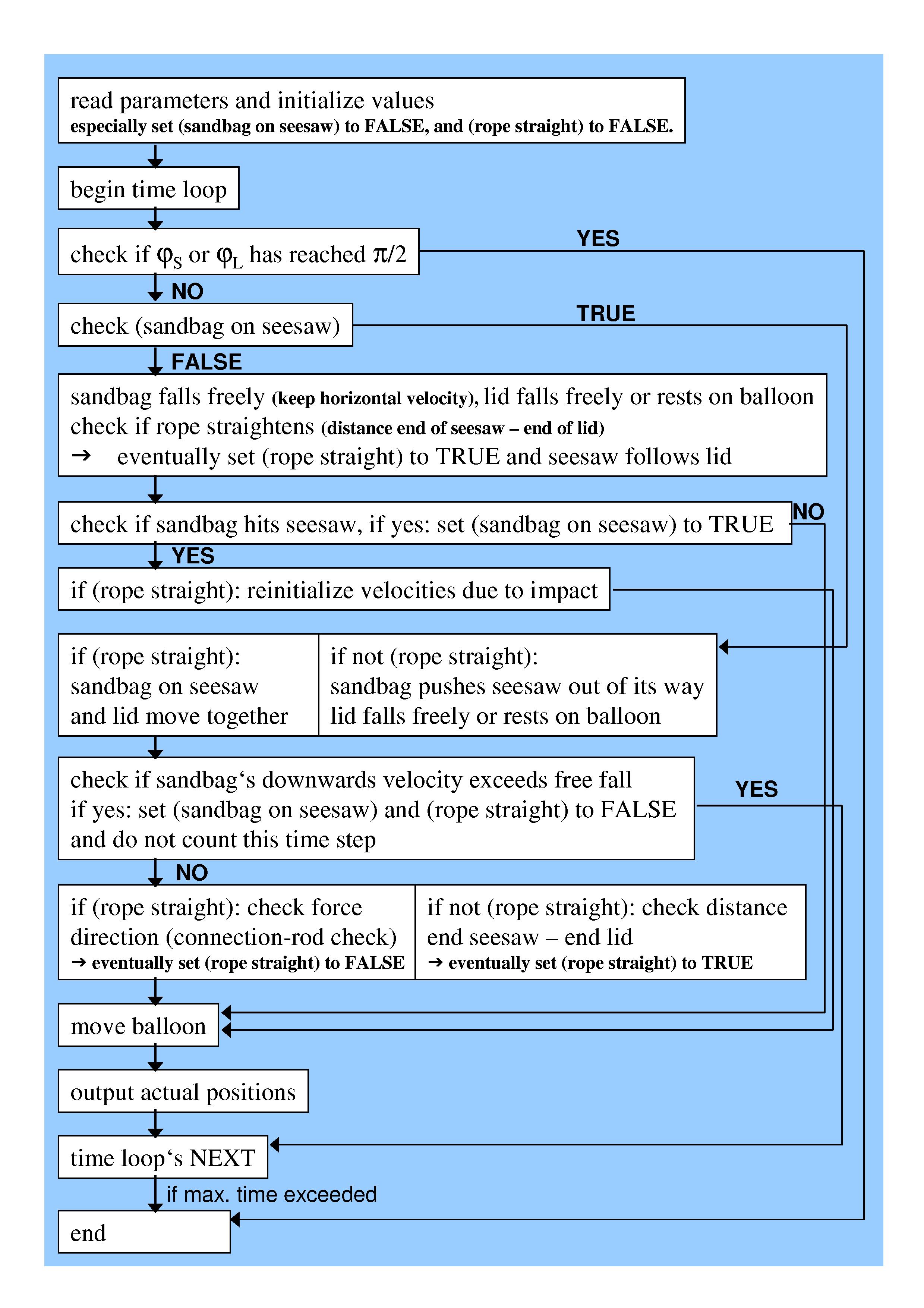}
\caption{Flow diagram for programming.}
\end{center}\end{figure}

One further difficulty, which shall be called the ''connecting-rod problem'',
has so far been overlooked. Of course, this will depend on the parameters,
that is weights, lengths and initial angles. With the weight of the lid pulling
down, one naturally expects the angular motion to slow down, that is
$\ddot{\varphi}_S$ and $\ddot{\varphi}_L$ become negative.
The direction of $\frac{d}{dt}(\frac{m_Sr_S^2\dot{\varphi}_S}{r_H})$ in
Fig.6 because of the second contribution besides the one with $\ddot{\varphi}_S$
in it, $2m_Sr_S\dot{r}_S\dot{\varphi}_S$, will not necessarily be reversed.
Furthermore, with directions of forces at the end of the seesaw as in Fig.6
there is no reason why we should not have $\frac{d}{dt}(\frac{m_Sr_S^2\dot{
\varphi}_S}{r_H})>\frac{m_Sgr_S\cos\varphi_S}{r_H}$. Then Fig.6 and (10) imply
that $\ddot{\varphi}_L<0$ and $\frac{I_L\vert \ddot{\varphi}_L \vert}{2r_L}
>\frac{m_L}{2}g\cos\varphi_L$. (We shall not allow somersaults, that is we
only regard $\varphi_S<\pi /2$ and $\varphi_L<\pi /2$. At $\varphi_S=\pi /2$
the sandbag would drop from the seesaw anyway. And the lid of the box is
assumed to engage when it reaches the vertical position. The projection factors
$\sin\alpha$ and $\sin(\alpha+\varphi_S-\varphi_L)$ are always positive;
the rope cannot come below the lid or above the right side of the seesaw.)
A constellation with $\frac{I_L\vert \ddot{\varphi}_L \vert}{2r_L}>
\frac{m_L}{2}g\cos\varphi_L$ would mean that, instead of tearing, forces would
be pushing into the rope from both ends. If then we solved (3) and (4) assuming
the distance between the right end of the seesaw and the end of the lid
constant, namely the length $l_R$, we would take the rope for a connecting
rod and say that the lid having aquired enough angular momentum could then
push against the seesaw from below via the ''rope''. In this case, in fact,
the rope will loosen. The motion of the lid and the sandbag has to be separated,
each merely further accelerated by the corresponding gravitational force
(our massless seesaw giving way to the sandbag), until again the rope
possibly straightens. This is why the general case of a horizontal velocity
of the sandbag and the lid already moving had been treated at impact.
The flow-diagram for integrating the equations of motion step by step in time
in a numerical simulation is given in Fig.7.

\section{Examples}

Four examples of calculations shall now be given. For each, parameters and
starting values as defined in section 2 are given.
$dt$ is the time step for the calculation. The balloon is in contact with
the lid of the box at the beginning.

\vspace{0.2cm}

\underline{first example:} Parameters:
$m_S$ = 0.6 kg, $r_H$ = 0.6 m, $l_R$ = 1.0 m, $r_L$ = 0.2 m,
\newline \phantom{m}\hspace{0.5cm}
$m_L$ = 0.5 kg, $x_{box}-x_{seesaw}$ = 0.73 m, $h_{seesaw}-h_{box}$ = 0.33 m,
$dt$ = 0.001 s.\newline \phantom{m}\hspace{0.5cm}
Starting values: $x_S$ = -0.3 m, $h_S$ = 1.0 m,
$\varphi_S$ = -25$^{\sf o}$, $\varphi_L$ = 5$^{\sf o}$.

\vspace{0.2cm}

Besides the horizontal and the vertical position of the sandbag in time,
we have also plotted its trajectory in space. Seesaw and lid angles over time
are united in one plot. The balloon's height as a function of time is drawn
together with the height limit imposed by the lid.
$t$=0, of course, is the time the sandbag starts falling. At $t$=0.42s it
hits the seesaw. However, the rope is not straight yet, thus the seesaw just
gives way to the falling sandbag. It already turns, such that at
$t$=0.51s when the rope straigtens and the lid begins to be pulled up,
the seesaw ''starting'' angle has changed from the initially given value.
However, here, immediately after the lid is pulled up for the first time, the
{\it connection-rod check} returns {\it true}, the rope loosens never to
become straight again in this example. The most complicated differential
equation (4) is only needed for a single time-step in this calculation!
The one impact has given the lid enough angular momentum. It reaches the
vertical position ($\varphi_L$=$\pi/2$) at $t$=0.82s.

\phantom{m}\hspace{-1.5cm}
\includegraphics[width=4cm,angle=270]{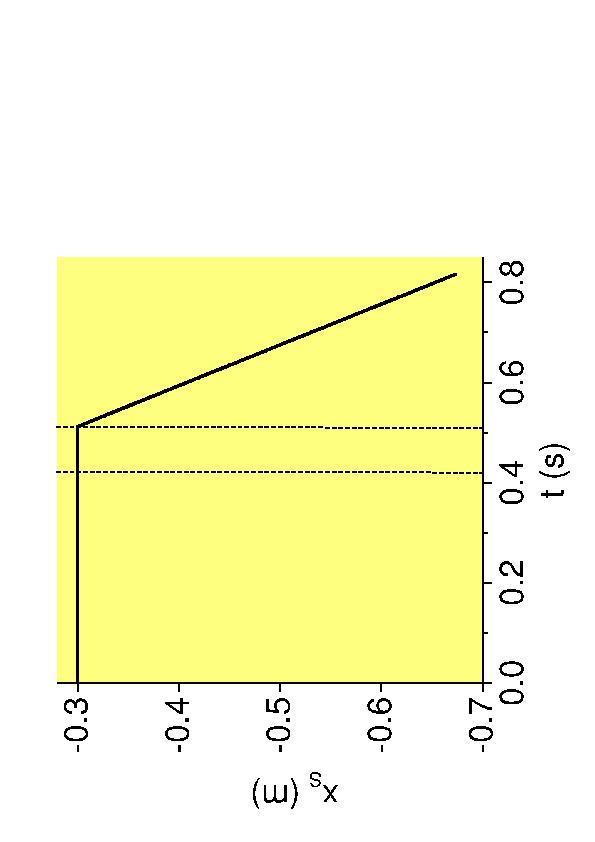}
\includegraphics[width=4cm,angle=270]{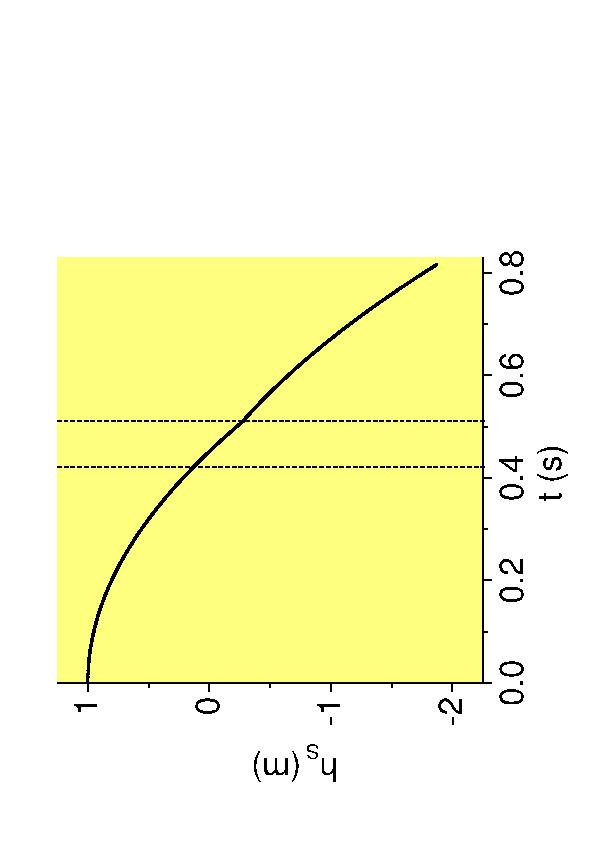}
\includegraphics[width=4cm,angle=270]{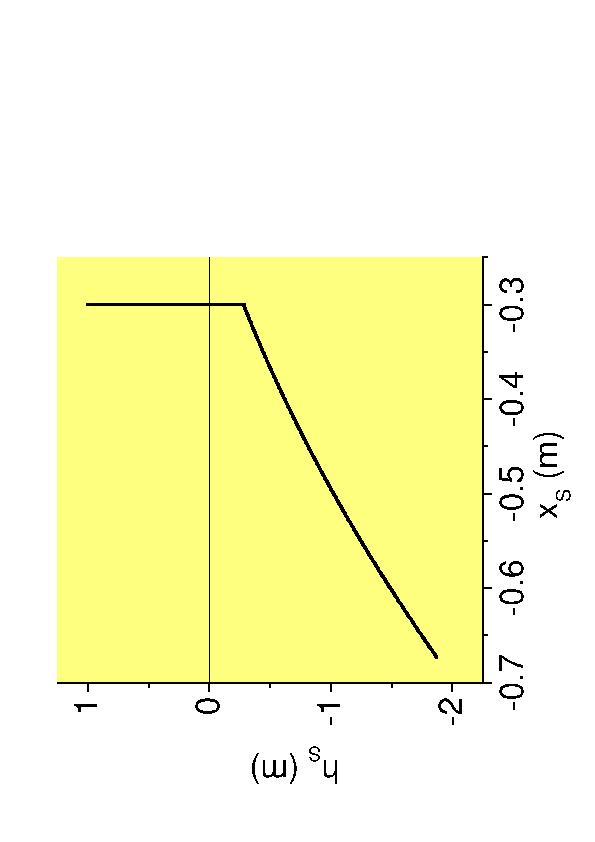}

\phantom{m}\hspace{-2cm}
\includegraphics[width=6.5cm,angle=270]{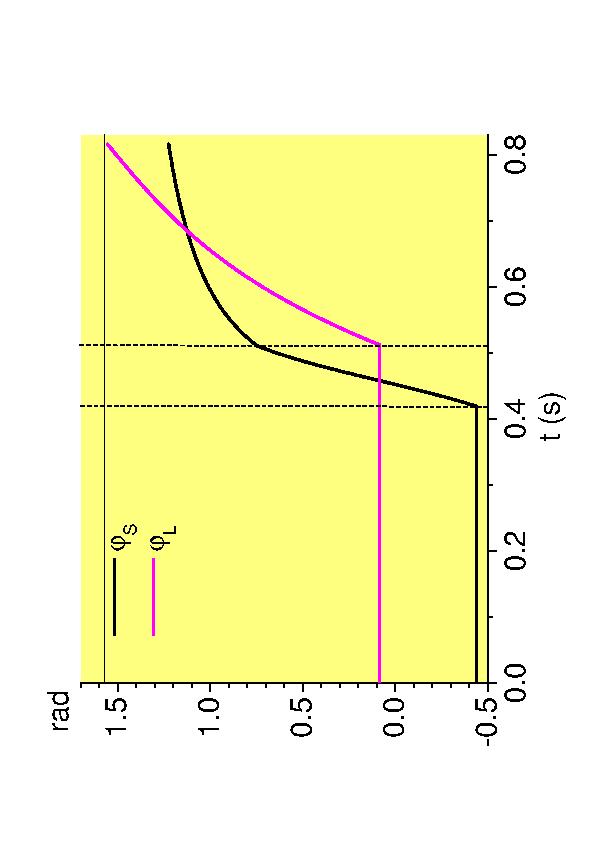}
\includegraphics[width=6.5cm,angle=270]{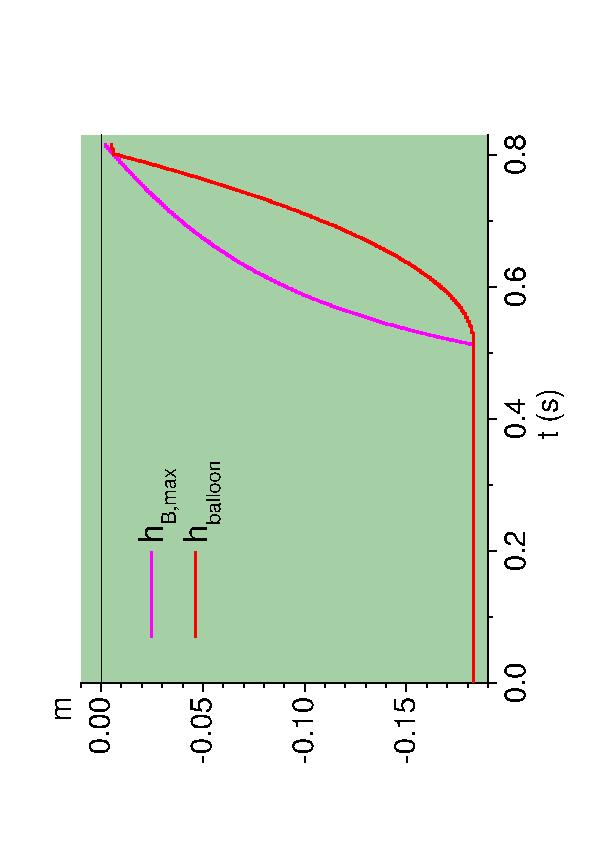}

\vspace{0.1cm}

{\small {\bf Fig.8.} First example.}

\vspace{0.2cm}

Note that by this time the sandbag is on the seesaw's left arm 2 meters
($\sqrt{x_S^2+h_S^2}$)
from the center. Our program does not contain a check for the sandbag to
reach the end of the seesaw (which could be added without problems,
nevertheless). In this example at later times the lid moves independently,
and the sandbag dropping from the seesaw would not change that. The lid
opens faster than the balloon can rise. Only shortly before the box has
fully opened does the balloon touch the lid once again. With the lid in
vertical position, the balloon can then rise freely. The time it takes to
turn on the radio depends on the length of the balloon string and has not
been calculated any more. Remark that the interesting part of the motion
from the rope's first straightening until the lid's reaching vertical
position only lasts from $t$=0.51s to $t$=0.82s, that is 0.31 seconds.

\vspace{0.2cm}

\underline{second example:} As a second example a case has been sought
where the rope stays straight at least for some time
while pulling up the lid. Parameters
and starting values are the same as in the first example except
$l_R$ = 0.62 m, $h_S$ = 0.5 m and $\varphi_S$ = -35$^{\sf o}$. 
The sandbag meets the seesaw at $t$=0.24s, the lid starts moving at
$t$=0.33s and reaches the vertical position at $t$=0.54s, that is already
0.21s later. The balloon
cannot catch up with the lid during that time. At the end the sandbag
is 0.6m from the center on the left arm of the seesaw, which is equal to
the length of the right arm in this case.
From $t$=0.49s on the sandbag cannot follow the seesaw any more and
looses touch for more than one time intervall (the rope loosens).
We only show the two angles
$\varphi_S$ and $\varphi_L$ as functions of time, however, also compare
them to the results that would have been obtained had the impact been
calculated following conservation of the sum of angular momenta, energy
or linear momenta (see section 3). These would be that the lid reaches
the vertical position at $t$=0.50s, $t$=0.47s or $t$=0.34s, respectively.
The last one definitely has to be disregarded as unphysical, though
(see section 3).

\phantom{m}\hspace{-2cm}
\includegraphics[width=6.5cm,angle=270]{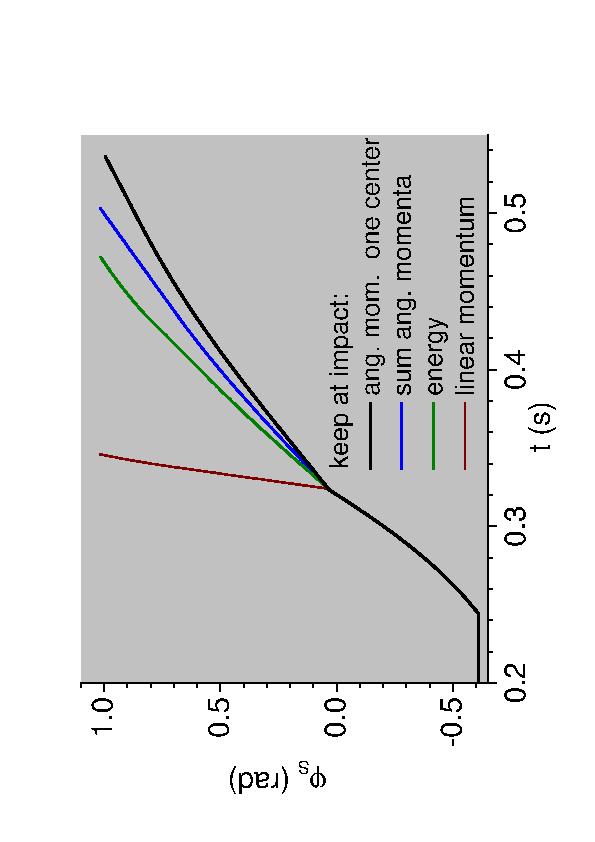}
\includegraphics[width=6.5cm,angle=270]{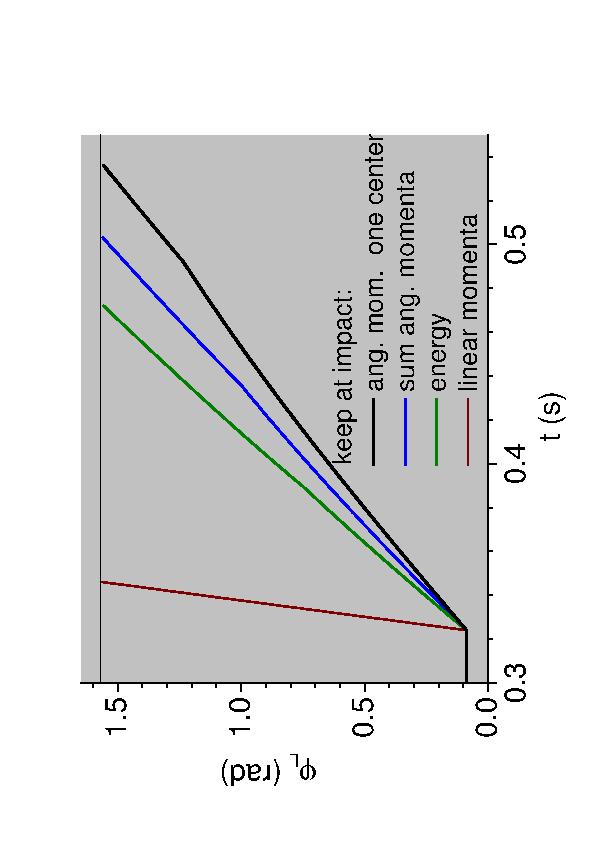}

\vspace{0.1cm}

{\small {\bf Fig.9.} Second example.}

\vspace{0.5cm}

\underline{third example:} Parameters and starting values are the same as in
the first example, except $dt$ = 0.00025 s and at the beginning
$h_S$ = 0.0 m, $\varphi_S$ = 0$^{\sf o}$, that is the seesaw is horizontal
and the sandbag is just placed on it, not let drop from a height.

\phantom{m}\hspace{-2cm}
\includegraphics[width=6.5cm,angle=270]{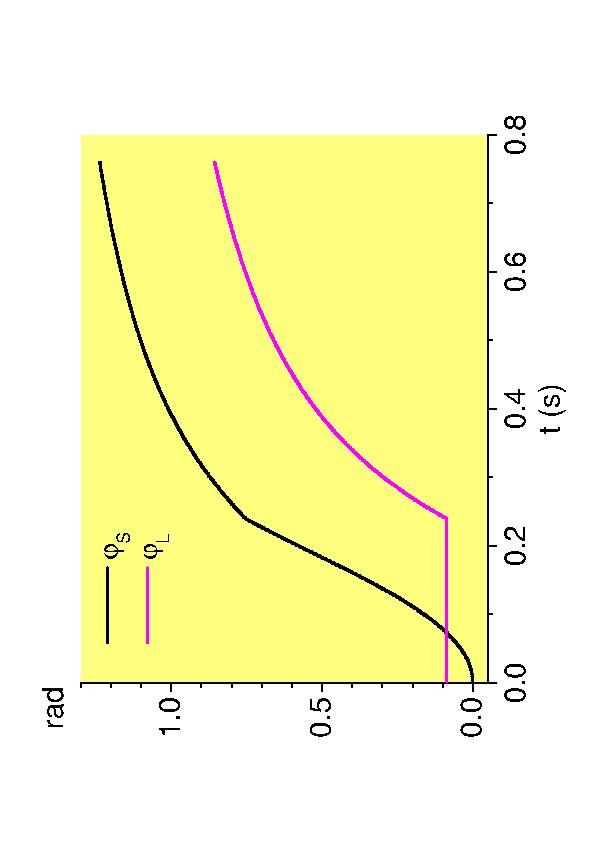}
\includegraphics[width=6.5cm,angle=270]{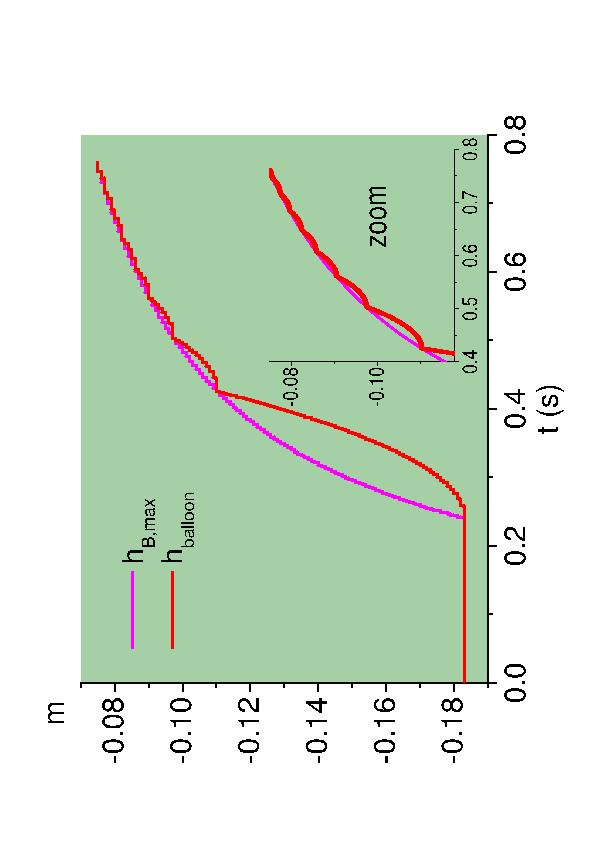}

\vspace{0.1cm}

{\small {\bf Fig.10.} Third example.}

\vspace{0.5cm}

This example has been chosen, because the lid opens so slowly that the
balloon catches up with it several times until practically staying trapped
touching the lid. Unfortunately, the chain reaction will not come to
freeing the ballon and switching on the radio in this case. The seesaw will
reach its vertical position at $t$=20s. However, there we 
assumed that the sandbag can slide an endless way on it, $r_S$ will be 1949m!
The lid will be at $\varphi_L$=74$^{\sf o}$. At $\varphi_S$=$\pi$/2 the
sandbag finally dropping from the seesaw and the rope loosening, even if one
persued this situation which has grown out of proportions, one constitutes
that the lid does not have sufficient angular velocity to continue up to
vertical position, but will quickly drop back down.

\vspace{0.5cm}

\underline{forth example:} Parameters:
$m_S$ = 0.7 kg, $r_H$ = 0.45 m, $l_R$ = 0.9 m, $r_L$ = 0.2 m,
\newline \phantom{m}\hspace{0.5cm}
$m_L$ = 0.4 kg, $x_{box}-x_{seesaw}$ = 0.64 m, $h_{seesaw}-h_{box}$ = 0.29 m,
$dt$ = 0.001 s.\newline \phantom{m}\hspace{0.5cm}
Starting values: $x_S$ = -0.3 m, $h_S$ = 0.5 m,
$\varphi_S$ = -20$^{\sf o}$, $\varphi_L$ = 5$^{\sf o}$.

\phantom{m}\hspace{-2cm}
\parbox{10cm}{\includegraphics[width=6.5cm,angle=270]{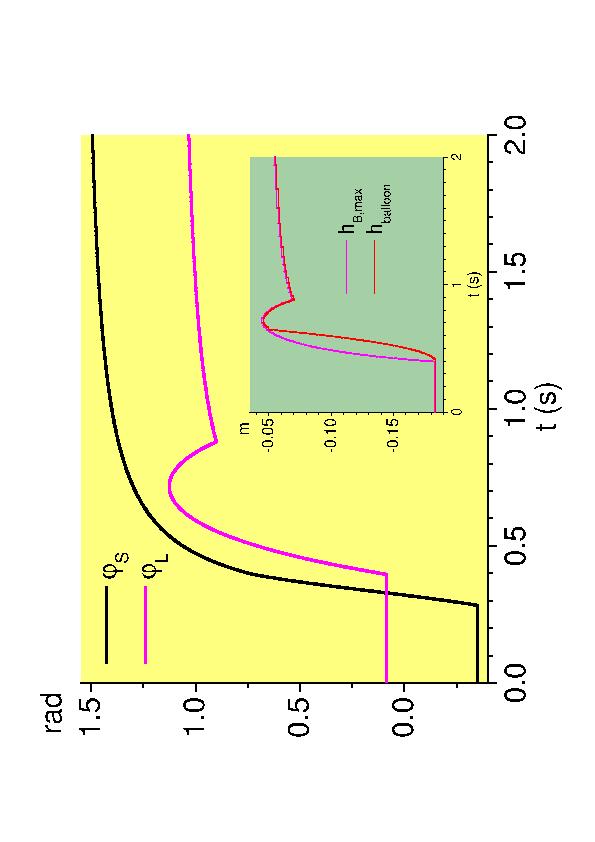}

\vspace{0.1cm}

\phantom{m}\hspace{2cm}
{\small {\bf Fig.11.} Forth example.}}
\parbox{7cm}{
Whereas in the third example the rope stayed straight after the first impact
at $t$=0.33s, here at $t$=0.40s the lid gets such a hard hit that immediately
afterwards the rope loosens again (connection-rod condition). The lid jumps
up, falls back and only when its fall has again straightened the rope ($t$=0.88s)
is pulled up by the counterweight of the sandbag.
However, as in the last
example, the seesaw will reach vertical position at about $t$=10s
($r_S$ will be 344m!), whereas the lid never will.
}

\vspace{0.4cm}

The balloon catches up with the lid at $t$=0.64s, to
be pressed down by it again and then follow it up again staying ''glued''
underneath it. 

\section{Conclusions}

In conclusion Kermit's What-Happens-Next-Machine has been treated as an
analytical mechanics and programming exercise.
The two lever arms of the seesaw and the box's
lid tied together by the rope can serve as a quite challenging example for
finding the equation of motion using
the Lagrange formalism. The most difficult point,
however, is the decision which quantities are conserved and which are not
at impact when the first force on the rope appears.
A simulation has been set up and examples with different behaviour have
been shown. It has been found that in cases where the counterweight of the
falling sandbag is able to fully open the box to free the balloon,
the interesting phase of the procedure only lasts some tenths of a second
(this cannot be known from the original scene from Sesame Street, because
there the machine does not work properly). Some simplifications have been
made to be able to solve the problem theoretically (for example: no friction,
neglection of the seesaw's mass). At the stage of development,
including a mass for the seesaw would mean merely a small extension for who
wants to refine the simulation. Attention has to be paid to the case that
the seesaw then contributes inertia even if the sandbag is not in touch with it.
As has been appreciated by some of our students, in the context of teaching
theoretical mechanics, the
What-Happens-Next-Machine can be given as an amusing tricky problem.

\section*{Acknowledgements}
Thanks to Elke Scheer and Wolfgang Belzig, who held the second year's physics
lecture to which the What-Happens-Next-Machine was put as a supplementary
exercise in the form of a competition.
The students Christoph K\"olbl and Matthias Haas built the machine shown in
Fig.s 3 and 4 and Timm Treskatis' quite complete theoretical analysis drew my
attention to some points in which my own original calculation had to be
improved.

\section*{Attachments:}
exercise sheet from physics course (in German);
skit of the scene is in English, \newline 
program in FORTRAN, geometric angle and max. balloon height calculation

\section*{References and Annotations}

\centerline{--------------------------------------------}

\section*{Attachments}

\subsection*{Mutual dependence of the seesaw's and lid's angles
and maximum height of the balloon}

Instead of the positions $(x_{seesaw},h_{seesaw})$ and $(x_{box},h_{box})$
for the center of the seesaw
and the hinge of the lid for our programs parameters $f$ and $\gamma$ are
chosen, that is the distance between these two points and the angle the line
between them makes with the horizontal. $d$, $\sigma$, $\tau$ and $\alpha$
are defined as can be seen from the drawing. Now the cosine- and sine-theorem
can be used to get $\varphi_L$ when given $\varphi_S$.

\begin{center}
\includegraphics[width=12cm,angle=270]{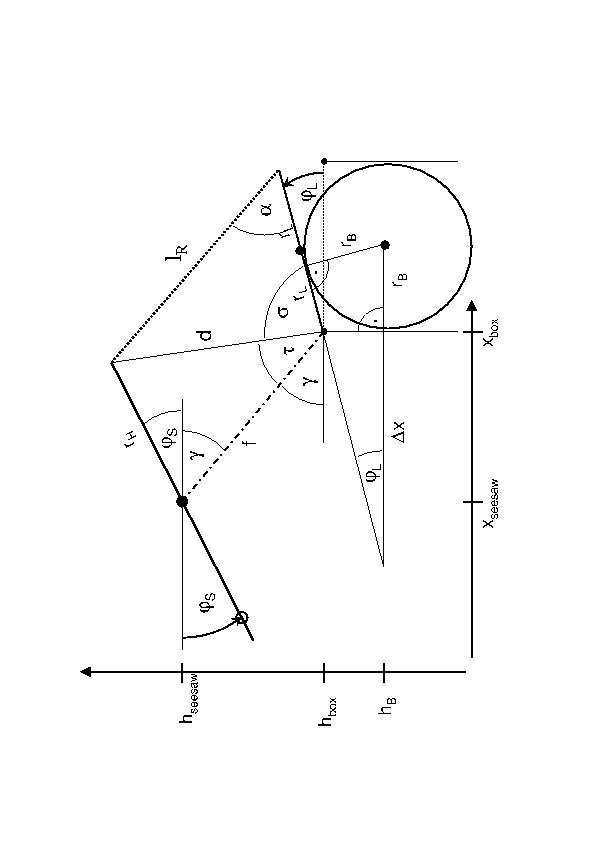}
\end{center}

$d=\sqrt{r_H^2+f^2-2r_Hf\cos (\varphi_S+\gamma)}$, \phantom{nix}
$\tau =\arcsin (r_H \;\sin (\varphi_S+\gamma)\; /d)$, \newline
$\sigma=\arccos ((4r_L^2+d^2-l_R^2)/(4r_Ld))$, \phantom{nix}
$\varphi_L=\pi-\gamma-\tau-\sigma$, \phantom{nix}
$\alpha=\arccos((4r_L^2+l_R^2-d^2)/(4r_Ll_R))$.\newline
The above calculation holds if the rope is straight.
Having expressed $d$ in terms of $\varphi_S$, the general constraint on
$\varphi_L$ reads \ 
$d^2+(2r_L)^2-2d\cdot (2r_L)\cdot \cos (\pi-\gamma-\tau-\varphi_L) \le l_R^2$.

As long as the lid has not come to vertical, the balloon must stay below it.
We here calculate the maximum possible height of its center. There are slightly
different ways to do that, leading to the same result, of course. One could
use the fact that the two drawn balloon radii together with the enclosed pieces
of the box's wall and lid form a symmetric kite with two right angles. Here
use the triangle with angle $\varphi_L$ formed to the left of the box by the
extensions of the lid and the horizontal balloon radius. Putting together
$\tan\varphi_L=(h_{box}-h_{B,max})/\Delta x$ and $\sin\varphi_L=r_B/(r_B+\Delta x)$
to eliminate $\Delta x$ one deduces
$r_B+(h_{box}-h_{B,max})/\tan\varphi_L=r_B/\sin\varphi_L$
and so the constraint on the balloon's motion is:
$h_B \le h_{box}-r_B\cdot (1-\sin\varphi_L)/\cos\varphi_L$.

\newpage

\subsection*{Program}

{\scriptsize \begin{verbatim}
      program kermit
      implicit none

      integer cr,sos,steps,count,straight
      double precision hs,xs,phiS,phiL,gamma,f,pi,rL,rB,rH,ms,mL,iL,rs
      double precision rhoair,rhohe,tmax,dt,g,vvs,phisp,rsp,rspp,alpha
      double precision xfs,hfs,xfl,hfl,xsold,t,philp,vhs
      double precision hb,vb,hbold,hbl,hbf,dphi,phisold,philold
      double precision phispp,d,arccos,hsold,vvsold,vhsold,rsold,rspold
      double precision philplus,philminus,numerator,denominator
      double precision philstart,airline,lR,deriv1,deriv2,phispold
      double precision pv,ph,coefa,coefb,coefc,coefd,philpold
      common pi,rH,rL,f,gamma,lR

      open(11,file='whatnext.inp',status='old')
      open(12,file='whatnext.dat',status='unknown')

************************************************************************
** This is what the input file 'whatnext.inp' should look like        **
************************************************************************
**                                                                     *
* 1.0 ; hs = initial height of sandbag in meters                       *
* -0.3 ; xs = initial horizontal position of sandbag in meters         *
* -25.0 ; phiS = initial angle of the seesaw in degrees                *
* 5.0 ; phiL = initial angle of the lid in degrees                     *
* 0.8 ; f = distance  center of seesaw - lid's hinge in meters         *
* 1.0 ; lR = length of the rope in meters                              *
* 27.0 ; gamma=angle of line seesaw center-lid hinge to horizontal(deg)*
* 0.2 ; rL = half length of the lid in meters                          *
* 0.6 ; rH = length of right side of seesaw in meters                  *
* 0.6 ; ms = mass of the sandbag in kilograms                          *
* 0.5 ; mL = mass of the lid in kilograms                              *
* 1.29 ; rhoair =  density of air in kilograms per cubic meter         *
* 0.18 ; rhohe = density of helium in kilograms per cubic meter        *
* 1.0 ; tmax = maximum time for calculation in second                  *
* 0.001 ; dt = time step in seconds                                    *
**                                                                    **
************************************************************************

      read(11,*) hsold
      read(11,*) xsold
      read(11,*) phiS
      read(11,*) phiL
      read(11,*) f
      read(11,*) lR
      read(11,*) gamma
      read(11,*) rL
      read(11,*) rH
      read(11,*) ms
      read(11,*) mL
      read(11,*) rhoair
      read(11,*) rhohe
      read(11,*) tmax
      read(11,*) dt
      pi=4.0*atan(1.0)
      phiS=phiS/180.0*pi
      phiL=phiL/180.0*pi
      gamma=gamma/180.0*pi
      phisold=phiS
      philold=phiL
      philstart=phiL
      rB=rL
      iL=4.0/3.0*mL*rL*rL
      g=9.81
      vvs=0.0
      vvsold=0.0
      vhs=0.0
      vhsold=0.0
      phisp=0.0
      philp=0.0
      phispold=0.0
      philpold=0.0
      vb=0.0
      phispp=0.0
      dphi=pi/200.0
      xfs=-f*cos(gamma)+rH*cos(phiS)
      hfs=f*sin(gamma)+rH*sin(phiS)
      xfl=2.0*rL*cos(phiL)
      hfl=2.0*rL*sin(phiL)
      airline=sqrt((xfs-xfl)**2+(hfs-hfl)**2)
      if (airline.gt.lR) then
       print*,'The rope is too short for these starting parameters.'
       print*,airline
       stop
      endif
      if (phis.lt.0.0) then
       if (hsold.lt.-xsold*tan(phis)) then
       print*,'sandbag must not start underneath seesaw'
       stop
       endif
      else
       if (hsold.lt.xsold*tan(phis)) then
        print*,'sandbag must not start underneath seesaw'
        stop
       endif
      endif
      sos=0
      straight=0
      steps=int(tmax/dt)
      t=0.0

      if (abs(phil).le.1.0D-10) then
      hbold=-rB
      else
      hbold=-(1.0-sin(phiL))/cos(phiL)*rB
      endif

      do count=1,steps
      t=t+dt
      cr=0

c stopping condition
      if ((phiL.gt.pi/2.0-dphi).or.(phiS.gt.pi/2.0-dphi)) goto 90

c check if sandbag is on seesaw
      if (sos.eq.1) goto 20

c sandbag and lid move separately
      vvs=vvsold+g*dt
      vhs=vhsold
      hs=hsold-vvsold*dt-0.5*g*dt*dt
      xs=xsold+vhsold*dt
      phiL=phiL+philp*dt
      philp=philp-mL*g*rL/iL*cos(phiL)*dt
      if (phiL.lt.philstart) then
       phiL=philstart
       philp=0.0
      endif
      if (straight.eq.0) then
       phiS=phiS+phisp*dt
       xfs=-f*cos(gamma)+rH*cos(phiS)
       hfs=f*sin(gamma)+rH*sin(phiS)
       xfl=2.0*rL*cos(phiL)
       hfl=2.0*rL*sin(phiL)
       airline=sqrt((xfs-xfl)**2+(hfs-hfl)**2)
       if (airline.ge.lR) straight=1
      endif
      if (straight.eq.1) then
       call machphis(phiL,phiS)
       phisp=(phiS-phisold)/dt
      endif

c check if sandbag has hit seesaw
      if (atan(hs/xs).lt.phiS) goto 30

c initialize velocities at impact
      sos=1
      rs=-xs/cos(phiS)
      rsp=-vhs*cos(phiS)+vvs*sin(phiS)
      if (straight.eq.1) then
c keep angular momentum referred to ONE center
       d=sqrt(rH*rH+f*f-2.0*rH*f*cos(phiS+gamma))
       alpha=arccos((4.0*rL*rL+lR*lR-d*d)/(4.0*rL*lR))
       call machphiL(phiS,phiL)
       call machphiL(phiS+dphi,phiLplus)
       call machphiL(phiS-dphi,phiLminus)
       deriv1=(phiLplus-phiLminus)/2.0/dphi
       numerator=ms*rs*(vvs*cos(phiS)+vhs*sin(phiS))+(iL*philp*rH*
     1  sin(alpha+phiS-phiL))/(2.0*rL*sin(alpha))
       denominator=ms*rS*rS+(iL*rH*sin(alpha+phiS-phiL)*deriv1)/
     1  (2.0*rL*sin(alpha))
       phisp=numerator/denominator
       philp=deriv1*phisp
cc if we kept sum of angular momenta
cc sandbag around center of seesaw and lid around its hinge
c       call machphiL(phiS,phiL)
c       call machphiL(phiS+dphi,phiLplus)
c       call machphiL(phiS-dphi,phiLminus)
c       deriv1=(phiLplus-phiLminus)/2.0/dphi
c       phisp=(ms*rS*(vvs*cos(phiS)+vhs*sin(phiS))+iL*philp)/
c     1  (ms*rS*rS+iL*deriv1)
c       philp=deriv1*phisp
cc if we kept kinetic energy
c       call machphiL(phiS,phiL)
c       call machphiL(phiS+dphi,philplus)
c       call machphiL(phiS-dphi,philminus)
c       deriv1=(philplus-philminus)/2.0/dphi
c       numerator=ms*(vvs*cos(phiS)+vhs*sin(phiS))**2+iL*philp*philp
c       denominator=ms*rS*rS+iL*deriv1*deriv1
c       phisp=sqrt(numerator/denominator)
c       philp=deriv1*phisp
cc if we kept linear momentum - results in nonsense
c       ph=ms*vhs-mL*rL*philp*sin(phiL)
c       pv=-ms*vvs+mL*rL*philp*cos(phiL)
c       call machphiL(phiS,phiL)
c       call machphiL(phiS+dphi,philplus)
c       call machphiL(phiS-dphi,philminus)
c       deriv1=(philplus-philminus)/2.0/dphi
c       coefa=-ms*cos(phiS)
c       coefb=ms*rS*sin(phiS)-mL*rL*sin(phiL)*deriv1
c       coefc=-ms*sin(phiS)
c       coefd=-ms*rS*cos(phiS)+mL*rL*cos(phiL)*deriv1
c       rsp=(ph*coefd-pv*coefb)/(coefa*coefd-coefb*coefc)
c       phisp=(ph*coefc-pv*coefa)/(coefb*coefc-coefa*coefd)
c       philp=deriv1*phisp
      else
       phisp=vvs*cos(phiS)/rs+vhs*sin(phiS)/rs
      endif
c anyway, that's it for this time step
      goto 30

   20 continue
c sandbag is on seesaw
      if (straight.eq.1) then
      call machphiL(phiS,phiL)
      call machphiL(phiS+dphi,philplus)
      call machphiL(phiS-dphi,philminus)
      deriv1=(philplus-philminus)/2.0/dphi
      deriv2=(philplus-2.0*phiL+philminus)/dphi/dphi
      numerator=ms*g*rS*cos(phiS)-2.0*ms*rS*rsp*phisp-
     1 mL*g*rL*cos(phiL)*deriv1-iL*phisp*phisp*deriv2*deriv1
      denominator=ms*rS*rS+iL*deriv1*deriv1
      phispp=numerator/denominator
      else
      phispp=(g*cos(phiS)-2.0*rsp*phisp)/rS
      endif
      rspp=rs*phisp*phisp+g*sin(phiS)
      rsp=rsp+rspp*dt
      rS=rS+rsp*dt
      phisp=phisp+phispp*dt
      philp=deriv1*phisp
      phiS=phiS+phisp*dt
      xs=-rS*cos(phiS)
      hs=-rS*sin(phiS)
      vvs=rsp*sin(phiS)+rS*cos(phiS)*phisp
      vhs=-rsp*cos(phiS)+rS*sin(phiS)*phisp
      if (straight.eq.1) then
      call machphiL(phiS,phiL)
      else
      phiLp=philp-mL*g*rL/iL*cos(phiL)*dt
      phiL=phiL+philp*dt
      endif
      if (phiL.lt.philstart) then
       phiL=philstart
       philp=0.0
      endif
      
c check if sandbag falls faster than allowed by free fall
      if (vvs-vvsold.gt.g*dt) then
       sos=0
       straight=0
       t=t-dt
       hs=hsold
       xs=xsold
       vhs=vhsold
       vvs=vvsold
       rs=rsold
       rsp=rspold
       phiS=phisold
       phiL=philold
       phisp=phispold
       philp=philpold
       goto 100
      endif

      if (straight.eq.1) then
c connection rod check, check whether rope loosens
       if (ms*rS*rS*phispp+2.0*ms*rs*rsp*phisp.gt.
     1 ms*g*rS*cos(phiS)) then
        cr=1
        straight=0 
       endif
      else
c distance check, check whether rope straightens
c       print*,'check distance'
       xfs=-f*cos(gamma)+rH*cos(phiS)
       hfs=f*sin(gamma)+rH*sin(phiS)
       xfl=2.0*rL*cos(phiL)
       hfl=2.0*rL*sin(phiL)
       airline=sqrt((xfs-xfl)**2+(hfs-hfl)**2)
       if (airline.ge.lR) then
       straight=1
       endif
      endif

   30 continue
c move the balloon
      hbl=-(1.0-sin(phiL))/cos(phiL)*rB
      hbf=hbold+vb*dt+0.5*g*(rhoair-rhohe)/rhoair*dt*dt
      if (hbl.lt.hbf) then
      hb=hbl
      vb=0.0
      else
      hb=hbf
      vb=(hb-hbold)/dt
      endif

c update values
      hbold=hb
      xsold=xs
      hsold=hs
      vvsold=vvs
      vhsold=vhs
      rsold=rs
      rspold=rsp
      phisold=phiS
      philold=phiL
      phispold=phisp
      philpold=philp

c output
      write(12,123) t,xs,hs,phiS,phiL,hbl,hb,dble(cr),dble(straight)
  123 format (9F8.3)

  100 enddo

      print*,'time out'
      goto 95

   90 print*,'reached vertical position, t',t

   95 close(11)
      close(12)
      end

      double precision function arccos(x)
      implicit none
      double precision x,pi,tangens
      pi=4.0*atan(1.0)
      if (abs(x).le.1.0D-10) then
       arccos=pi/2.0
      else
       tangens=sqrt(1.0-x*x)/x
       arccos=atan(tangens)
       if (x.lt.0.0) arccos=arccos+pi
      endif
      return
      end

      subroutine machphiS(phiL,phiS)
      implicit none
      double precision phiS,phiL,pi,lR,rH,rL,f,gamma,dhere,tausigma
      double precision arccos,downangle,upangle
      common pi,rH,rL,f,gamma,lR
      tausigma=pi-phiL-gamma
      dhere=sqrt(f*f+4.0*rL*rL-4.0*f*rL*cos(tausigma))
      downangle=arccos((dhere*dhere+f*f-4.0*rL*rL)/2.0/dhere/f)
      upangle=arccos((dhere*dhere+rH*rH-lR*lR)/2.0/dhere/rH)
      phiS=downangle+upangle-gamma
      return
      end

      subroutine machphiL(phiS,phiL)
      implicit none
      double precision pi,rH,rL,gamma,lR,f,phiL,phiS,arccos,tau,sigma,d
      common pi,rH,rL,f,gamma,lR
      d=sqrt(rH*rH+f*f-2.0*rH*f*cos(phiS+gamma))
      tau=arccos((f*f+d*d-rH*rH)/2.0/f/d)
      sigma=arccos((4.0*rL*rL+d*d-lR*lR)/4.0/rL/d)
      phiL=pi-gamma-tau-sigma
      return
      end
\end{verbatim}}

\vspace{2cm}

\subsection{Skit of the scene from Sesame Street}

{\it (Kermit is standing next to a rope on a pulley that's attached to a sandbag.
The sandbag is suspended over a seesaw. The other end of the seesaw, pointing
down, has a string on it attached to the closed lid of a box containing an
inflated helium balloon. Its string is trailing out of the box and is tied to
the switch of a radio.)}

Kermit {\it (who has a pair of scissors)}: First of all, I'll cut the rope.
Then the sandbag will fall on the seesaw. Then that end of the seesaw will
go down, and the other end of the seesaw will go up.
...and the balloon will float up in the air! And you will notice that
the balloon is tied to the switch of the radio.

Kermit {\it (going over to the radio)}: You see, in the old days I had to walk
all the way over here to turn on my radio. It would tire my flippers.
It would waste time.
But not anymore ... thanks to the magic of What-Happens-Next! ...
First I cut the rope!
{\it (He does ... and the sandbag stays suspended in mid-air.)}

Kermit: And what happens next is ... the sandbag seems to be stuck!
The rope must be tangled ... {\it (He gives it a few good whacks,
but it still doesn't fall.)}
Oh, well, we'll just skip the sandbag part and go on to the seesaw.
I'll just give it a push. {\it (He does, but it won't budge.)}
The seesaw seems to be stuck ... {\it 
(He gives a few determined grunts and shoves, but to no avail.)}

Kermit: Oh, well, we'll just ignore the stuck seesaw and move to the box.
And what do you know, I'll bet the lid won't open.
{\it (He gets it open on the first try, though.)} Oh, there.
{\it (The balloon refuses to rise.)}
I'll give it a few good kicks! {\it
(He does, but the balloon stays put. By now he's really upset.)}
Come on, balloon! Are you going to float up in the air or not? {\it
(Kicks it again.)}
All right, balloon, I'm giving you one last chance! You float up in that
air right now, or I won't be responsible for what happens next! {\it
(He gives the box a vicious kick, and finally the balloon floats up in the
air, turning on the radio, which starts playing music.)}

Kermit: Thanks to the magic of What-Happens-Next. 
Once I get all the bugs out. {\it
(Just then, he sees the balloon carrying the radio away.)}

Kermit: What happens next is I gotta get another radio! {\it
(He jumps as the sandbag suddenly falls on the seesaw, which breaks in two.)}

\vspace{0.2cm}

[adapted from SilveryShoe@aol.com]


\begin{thebibliography}{6}
\bibitem{praxis}
In pratice the volume of helium has to be large enough to
lift up the balloon with its skin which in turn requires a large enough box.
The one in Fig.s 3 and 4 of about 25cm$\times$25cm proved too small for that
reason.
\end{thebibliography}
\end{document}